# High-Speed Multifunctional Photonic Memory on a Foundry-Processed Photonic Platform

Sadra Rahimi Kari[1], Marcus Tamura[2], Zhimu Guo[2], Yi-Siou Huang[3,4], Hongyi Sun[3,4], Chuanyu Lian[3,4], Nicholas Nobile[1], John Erickson[1], Maryam Moridsadat[2], Carlos A. Ríos Ocampo[3,4], Bhavin J Shastri[2,*], Nathan Youngblood[1,*]

[1]*Department of Electrical and Computer Engineering, University of Pittsburgh, Pittsburgh, PA 15261, USA*
[2]*Centre for Nanophotonics, Department of Physics, Engineering Physics & Astronomy, Queen's University, Kingston, ON, K7L 3N6, Canada*
[3]*Department of Materials Science & Engineering, University of Maryland, College Park, MD, 20742, USA*
[4]*Institute for Research in Electronics and Applied Physics, University of Maryland, College Park, MD, 20742, USA*
*Email: nathan.youngblood@pitt.edu, shastri@ieee.org

## Abstract

The integration of computing with memory is essential for distributed, massively parallel, and adaptive architectures such as neural networks in artificial intelligence (AI). Accelerating AI can be achieved through photonic computing, but it requires nonvolatile photonic memory capable of rapid updates during on-chip training sessions or when new information becomes available during deployment. Phase-change materials (PCMs) are promising for providing compact, nonvolatile optical weighting; however, they face limitations in terms of bit precision, programming speed, and cycling endurance. Here, we propose a novel photonic memory cell that merges nonvolatile photonic weighting using PCMs with high-speed, volatile tuning enabled by an integrated PN junction. Our experiments demonstrate that the same PN modulator, fabricated via a foundry-compatible process, can achieve dual functionality. It supports coarse programmability for setting initial optical weights and facilitates high-speed fine-tuning to adjust these weights dynamically. The result showcases a 400-fold increase in volatile tuning speed and a 10,000-fold enhancement in efficiency. This multifunctional photonic memory with volatile and nonvolatile capabilities could significantly advance the performance and versatility of photonic memory cells, providing robust solutions for dynamic computing environments.

## Introduction

Recent advances in photonic materials, devices, and architectures have shown considerable promise for applications in artificial intelligence (AI) and neuromorphic computing[1,2]. One of the major motivations for photonics in the context of AI is the ability to multiplex information using many optical wavelengths and process it simultaneously (using wavelength division multiplexing) in an analog photonic processing unit at high speeds. This approach allows for parallel processing at high speeds, offering a viable alternative to traditional computing hardware. This is particularly beneficial for overcoming limitations such as energy consumption, computing density, and scalability that challenge current digital processors.

To maximize the potential of photonic computing, most photonic processing units are designed around a reconfigurable array of memory cells. These cells are programmed to implement a matrix of trained weights[3–7]. Matrix-vector multiplication is then achieved through parallel multiply-accumulate operations as a vector of optical inputs passes through the memory array and is read out by photodetectors at the output of the array[8]. While the optical input signals can be modulated at very high speeds with established electro-optic effects (free carrier, Kerr, Franz-Keldysh, etc.), the memory cells themselves are often much slower. This disparity between the memory cell and modulator is often due to a variety of competing requirements[9], such as low insertion loss, low static power consumption, small device footprint, high bit precision, and high dynamic range.

Furthermore, for practical scalability of the memory array, low static power consumption and small device footprint are critically important. This has spurred research into non-volatile optical mechanisms for storing information on-chip. Among various demonstrated[9–11] nonvolatile optical effects—e.g., mechanical[12], memristive[13,14], ferroelectric[15], magneto-optic[16], etc.—phase-change materials (PCMs) have emerged as particularly promising[17]. PCMs offer a large refractive index change ($\Delta n \sim 0.5–2$), compact footprint (1 ~ 10 μm), long-term retention (~10 years[18]), and multilevel storage capabilities (up to 7 bits per cell[19]). However, the major challenges facing these devices are: 1) their slow electronic reconfigurability of 50 ns ~ 100 μs; 2) power-hungry operation (10 nJ ~ 1 μJ); and 3) limited endurance of 1,000 ~ 10,000 cycles. It has been noted previously that photonic computing architectures leveraging phase-change memory can be significantly less power efficient and slower than their digital counterparts when programming energy and time are considered[20,21]. Here, we propose and demonstrate a multifunctional photonic memory cell that addresses these challenges by combining the benefits of nonvolatile phase-change materials with the efficient and high-speed tunability of foundry-processed modulators in micro-ring structure. Notably, this multifunctional cell utilizes a waveguide-integrated PN junction for both programming the PCM via localized heating under forward bias and high-speed fine-tuning of the optical weights under reverse bias conditions. Our co-integrated memory with both nonvolatile and volatile functionality aims to extend the domains of photonic computing beyond accelerating machine learning inference to fast and efficient in situ training and online learning.

## Results

Our device concept is shown in **Figure 1a,** where we have integrated two PN junctions into the left and right sides of a micro-ring resonator (MRR) with independent electrical control. This independent control allows us to selectively switch the side of the MRR containing the PCM to improve efficiency and localize heat generation. When an applied electrical pulse exceeds the threshold voltage of the PN junction (~0.7V), the PN junction is forward-biased and can locally heat the PCM, triggering a phase transition and enabling course tuning of the optical resonance. Fine-tuning of the resonance can be achieved through reverse biasing the PN junction, which changes the carrier concentration (and, therefore, effective index) in the waveguide. While this free carrier effect is volatile and relatively small, combining this high-speed fine-tuning with the slow-speed course-tuning of PCMs allows for both precise and efficient control over the optical weights. This is particularly useful in the context of on-chip training, where small changes in the optical weights can be rapidly implemented through the reverse-biased PN junction. Only when the accumulated weight change exceeds the tuning range of the PN junction is the PCM reprogrammed, thus improving the memory cell endurance and the energy efficiency and speed of the training process.

In **Figure 1b,** we plot the measured current density of a fabricated PN junction in both the forward and reverse bias conditions. Removing carriers from the waveguide by changing the depletion width results in a red shift of the resonance since for a silicon waveguide, $\Delta n_{eff} = -\alpha \Delta N - \beta \Delta P^{0.8}$, where $\Delta n_{eff}$ is the change in effective index of the waveguide, $\Delta N$ and $\Delta P$ are the changes in the electron and hole concentrations, and $\alpha$ and $\beta$ are constants which depend on the wavelength[22]. It is worth noting that volatile tuning of the resonance can also be achieved through forward biasing the junction between 0V and ~1.5V, causing a blue shift due to carrier injection. However, for forward biasing beyond the threshold voltage, heating of the waveguide and free carrier absorption begins to dominate, which reduces efficiency and extinction ratio. For large forward voltages (>4V), the heating can be significant enough to reach the crystallization and melting temperatures of the PCM, enabling reversible switching.

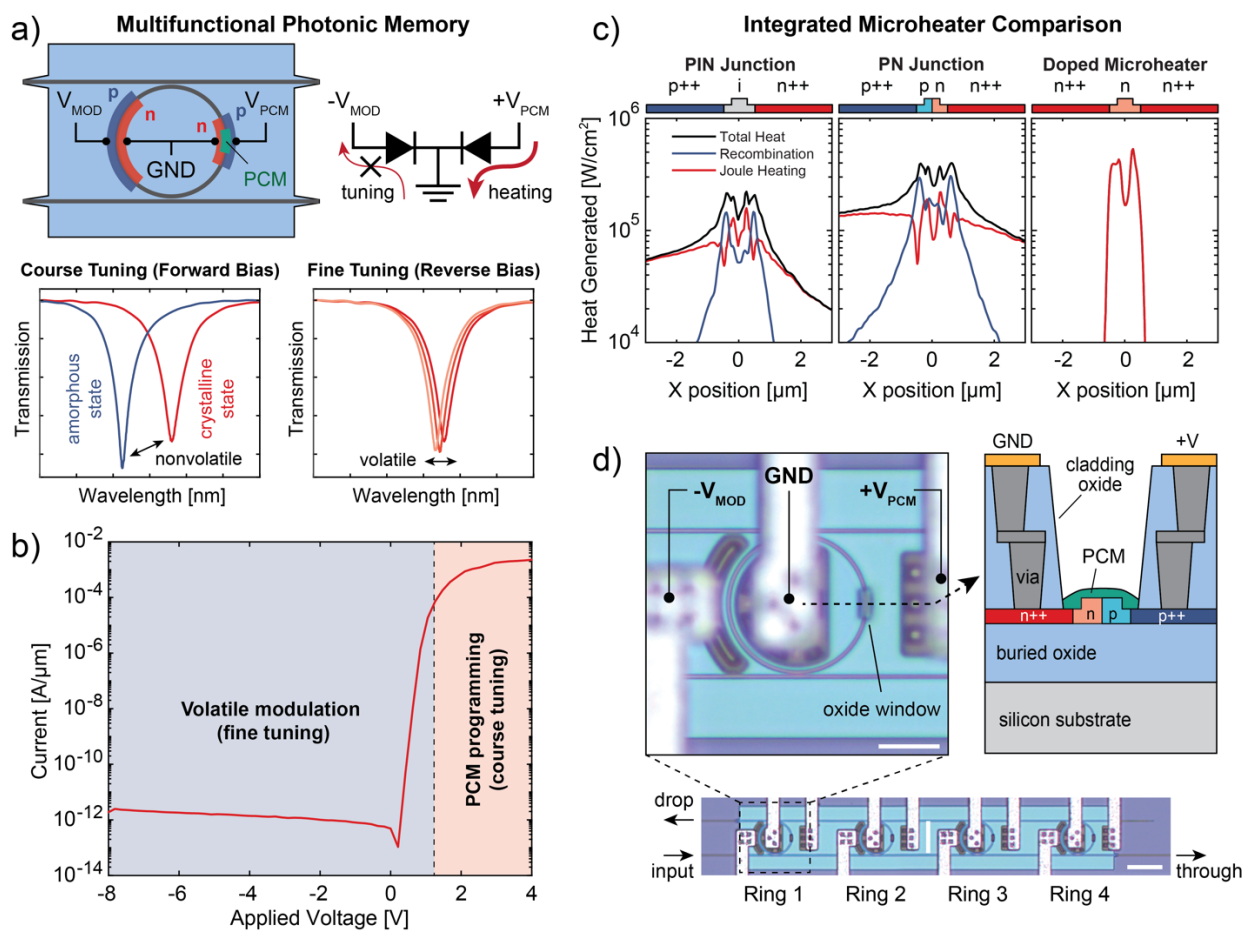

**Figure 1: Overview of multifunctional photonic memory concept. a)** Illustration and working mechanism of multifunctional photonic memory leveraging a PN junction microheater and modulator. Course nonvolatile tuning of the PCM is achieved by applying a forward-biased pulse to locally heat the waveguide while reverse biasing the PN junction modulates carriers in the depletion region for volatile fine-tuning. **b)** Measured current density as a function of voltage applied to the waveguide-integrated PN junction. Biasing the PN junction below ~1.5V enables volatile tuning without significant heating of the PCM. **c)** Modeling comparison of heat generated in PIN, PN, and n-doped waveguide-integrated microheaters. All simulations use a fixed +5V forward bias and 1 μm spacing between n++/p++ regions. **d)** Optical microscope image of a fabricated micro-ring array (bottom, 25 μm scale bars) and magnified image of a single micro-ring with integrated multifunctional memory cell (top left, 10 μm scale bar). Cross-sectional view of PN microheater with PCM deposited in oxide window (top right).

**Figure 1c** compares the simulated heat generation from three foundry-compatible microheater designs employing PIN, PN, and n-doped microheaters. These simulations were performed using COMSOL Multiphysics using the same material properties and methods we have explored in prior work[23]. For all designs, the separation between the heavily doped p++ and n++ regions was fixed at 1 μm (i.e., 1-μm-long intrinsic, PN, and n-doped regions), and +5 V was applied in the forward-biased direction. The observed peaks in the heat generation are mainly due to discontinuities in the silicon thickness (i.e., waveguide vs slab height) and changes in doping concentration and type. For the PIN and PN junctions, we observe heat generation due to both carrier recombination and Joule heating, while for the n-doped microheater, only Joule heating occurs due to the presence of only one carrier type.

In terms of heating efficiency, the n-doped heater maximizes the heat generated near the waveguide while minimizing heating in the heavily doped contacts compared to the PIN and PN microheaters. However, fine-tuning with an n-doped microheater will be dominated by Joule heating, which limits fine-tuning to sub-MHz speeds. Compared to the PIN microheater, the PN microheater generates more heat at the same forward bias voltage and can also be reverse biased to enable fine-tuning of the optical weights. While PIN and n-doped microheaters have been successfully used in the past to electrically switch PCMs[24–26], volatile tuning is only possible in the forward-biased direction[27]. This type of fine-tuning requires a constant current to be applied, which increases the static power dissipation of the device to several mW levels and reduces efficiency in the system. Additionally, for PIN junctions, the maximum tuning speed is limited by carrier recombination in the intrinsic region, resulting in slow (~15 MHz) volatile tuning speeds[27]. Using our PN junction design, we can achieve fine-tuning of the optical weights with sub-nanosecond speeds and with less than 80 pA of leakage current at a reverse bias of -8V, resulting in less than 640 pW of power dissipation per device.

Our multifunctional memory cell is particularly well suited for the broadcast and weight architecture[28], which requires accurate tuning of the resonance position for each optical weight[29–31]. Four multifunctional photonic memory cells sharing common bus waveguides for both positive and negative weighting are shown in **Figure 1d**. These devices were fabricated at Advanced Micro Foundry (AMF) using their active silicon photonic process (AMFSiP). Oxide windows etched down to the waveguide by AMF were used to integrate $Ge_2Sb_2Se_4Te_1$ (GSST) and $Sb_2Se_3$ after device fabrication (visible in the magnified image of a single memory cell in **Figure 1d**). Sputtering was then used to deposit 30 nm of either GSST or $Sb_2Se_3$, followed by a 30 nm oxide cladding to prevent oxidation (see Methods). A cross-sectional view of the final device is illustrated in **Figure 1d**.

To achieve reliable switching, all devices underwent a post-deposition anneal to ensure the PCM is initially in the crystalline state. The importance of this annealing step can be seen in **Figure 2**. In **Figure 2a**, we show the crystallization process of a memory cell with 5 μm of GSST without an initial hot plate anneal. The PN microheater was used to crystallize the GSST using 10 ms pulses of increasing amplitude ranging from 4V to 5V. As the pulse amplitude increases above the glass transition temperature of the GSST, the resonance experiences a blue shift while the extinction ratio decreases. However, the resonance eventually shifts toward longer wavelengths, as expected when the GSST is fully crystallized at higher pulse amplitudes. This blue shift can be explained by a topographical change in the distribution of GSST where material migrates away from the waveguide when heated above the glass transition temperature, as illustrated in **Figure 2a**, thus lowering $n_{eff}$ and shifting the resonance to shorter wavelengths. We confirm that this process is indeed topographical rather than ablation or oxidation by reversibly switching the device between the amorphous and crystalline states after initial crystallization (see **Figure 2b**). Amorphization was achieved using 100 µs pulses with an amplitude ranging from 7V to 8V. While the device can still function as nonvolatile memory, the reduced material above the waveguide after this reflow process lowers the maximum wavelength shift that can be achieved between the crystalline and amorphous states.

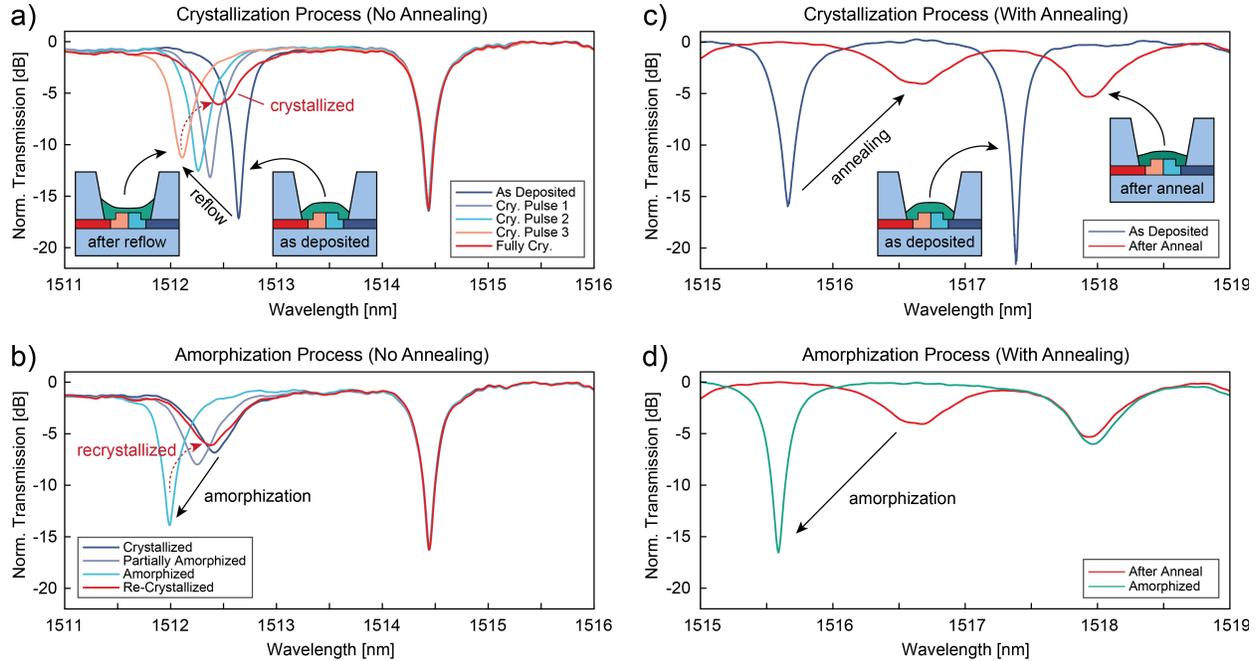

**Figure 2: Effects of annealing step on deposited PCM (Ge$_2$Sb$_2$Se$_1$Te$_4$). a)** Observed initial spectral blue shift prior to reaching the fully crystallized state due to reflow of the as-deposited PCM above the glass transition temperature. **b)** After reflow, reversible switching is achieved, demonstrating a topographical change in the PCM rather than a structural or chemical change. **c)** Observed spectral redshift of both resonances after annealing the entire chip on a hot plate. The large red shift is expected and indicates minimal reflow of the PCM after annealing. **d)** Spectrum of microring with re-amorphized PCM matches well with the initial as-deposited state, demonstrating minimal topographical changes after hot plate annealing.

To address this issue, we first perform a 200°C anneal on a hot plate for 10 minutes. The spectra before and after this anneal step can be seen in **Figure 2c** for another set of memory cells with the same geometry and length of GSST as in **Figure 2a-b**. When the GSST is crystallized, there is a clear shift to longer wavelengths and a reduced extinction ratio after annealing. This is expected for crystalline GSST since the refractive index and loss both increase relative to the as-deposited amorphous state. We observe minimal reflow in these devices, as evidenced by the ability to re-amorphize the GSST to a resonance position and extinction ratio very close to the initial as-deposited state (**Figure 2d**). While these results shown are for GSST, we observed the same behavior in Sb$_2$Se$_3$ samples with and without the post-deposition anneal step.

To demonstrate nonvolatile programming, we used memory cells with 2 µm of PCM (GSST and Sb$_2$Se$_3$) and set the optical weights using a series of amorphization pulses. **Figure 3a-b** shows the spectra of two memory cell arrays containing four MRRs, each before and after programming. In these experiments, Ring 1 was fully amorphized, Ring 4 was left in the crystalline state, and Rings 2 and 3 were set to an arbitrarily chosen partially crystalline state. While the resonance shift is slightly smaller for the devices with $Sb_2Se_3$, the lower loss results in a higher extinction ratio in the crystalline state (**Figure 3b**) compared to the rings with the same length of GSST (**Figure 3a**). This tuning range can be further increased by increasing the PCM length. While the loss begins to dominate for longer GSST lengths (see **Figure 2**), we observed reasonably high extinction ratios (>12 dB) in devices with 5 μm long $Sb_2Se_3$, which were able to achieve spectral shifts >800 pm.

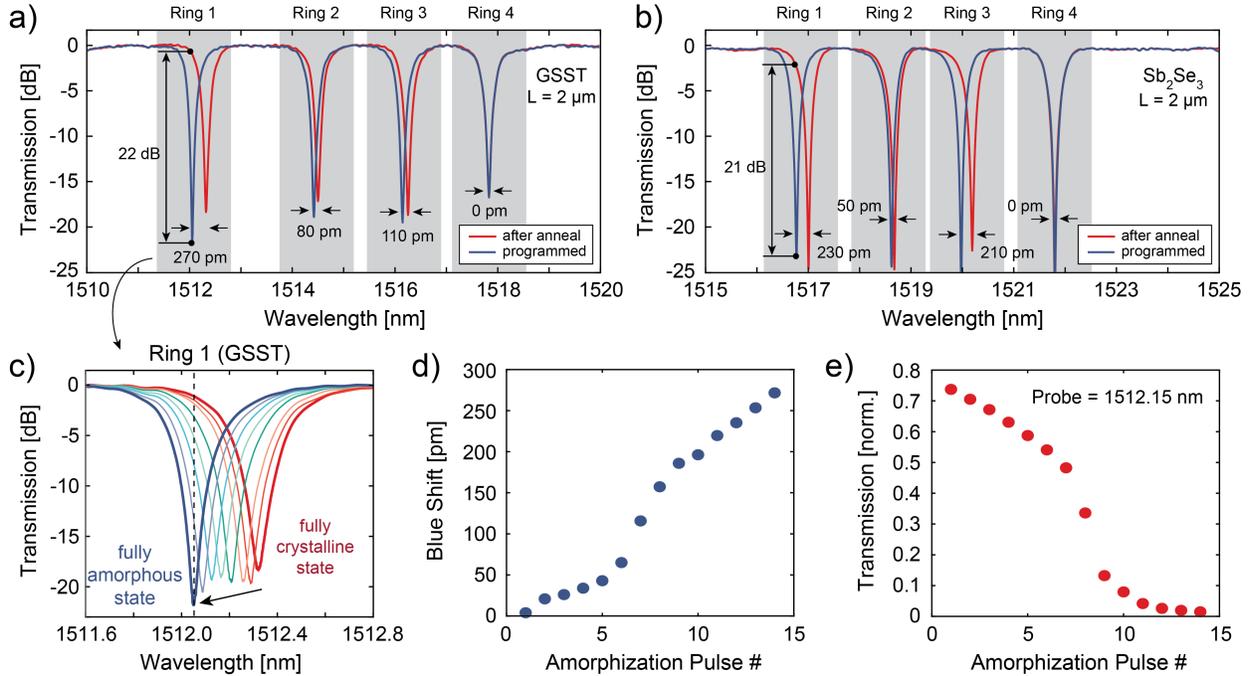

**Figure 3: Nonvolatile tuning of photonic memory. a-b)** Nonvolatile programming of arbitrary optical weights using memory cell arrays with four MRRs containing **a)** GSST and **b)** $Sb_2Se_3$. For these experiments, Ring 1 was fully amorphized, Rings 2 and 3 were partially amorphized, and Ring 4 was left in the crystalline state. **c)** Incremental amorphization of GSST using a PN microheater (probe wavelength indicated by dashed line). **d)** Measured blue shift and **e)** transmission state of Ring 1 from **c)** as a function of amorphization pulse number. Amorphization pulses were 100 μs in width with an increasing voltage ranging from 7V to 8V.

Coarse control of the optical weights is shown in **Figure 3c**, where amorphization pulses of increasing amplitude are used (100 μs pulse width with amplitude ranging from 7V to 8V). Here, we achieve 14 distinct levels (3.8 bits) and shift the resonance position by 270 pm (see **Figure 3d**). This large spectral shift allows us to achieve an insertion loss of -1.3 dB and an extinction ratio of 22 dB between the fully crystalline and amorphous states. In **Figure 3e**, we plot the transmission measured at a probe wavelength of 1512.15 nm (dashed line in **Figure 3c**) on a linear scale.

Having demonstrated coarse tuning of the optical weights, we also show volatile fine-tuning by biasing PN junctions below the threshold voltage. **Figure 4a** shows the measured resonance shift of the PN junction (19.2 μm arc length) under reverse bias conditions. As the reverse bias voltage increases, carriers are removed from the waveguide, and the depletion region increases. This results in a redshift of the spectra. Our measurements agree well with the analytical model found in Chrostowski et al.[22] for a PN modulator with n- and p-doping in the junction equal to $5 \times 10^{17}$

cm$^{-3}$. As the PN junction in our current devices only occupied 42% of the ring, this modulation efficiency could be further improved by a factor of ~2× simply by increasing the fraction of our memory cell covered by the PN junction.

It is also possible to achieve volatile tuning using a forward bias voltage. In this configuration, carriers are injected into the waveguide (rather than removed) and cause a blue shift of the resonance due to plasma dispersion. The change in carrier density under a forward bias voltage can be much larger than under a reverse bias, and therefore, the resonance shift is greater as well. In **Figure 4b**, we plot the measured shift in resonance for the case of the 10 μm PN microheater under forward bias. After the applied bias exceeds ~2× the threshold voltage of the PN junction,

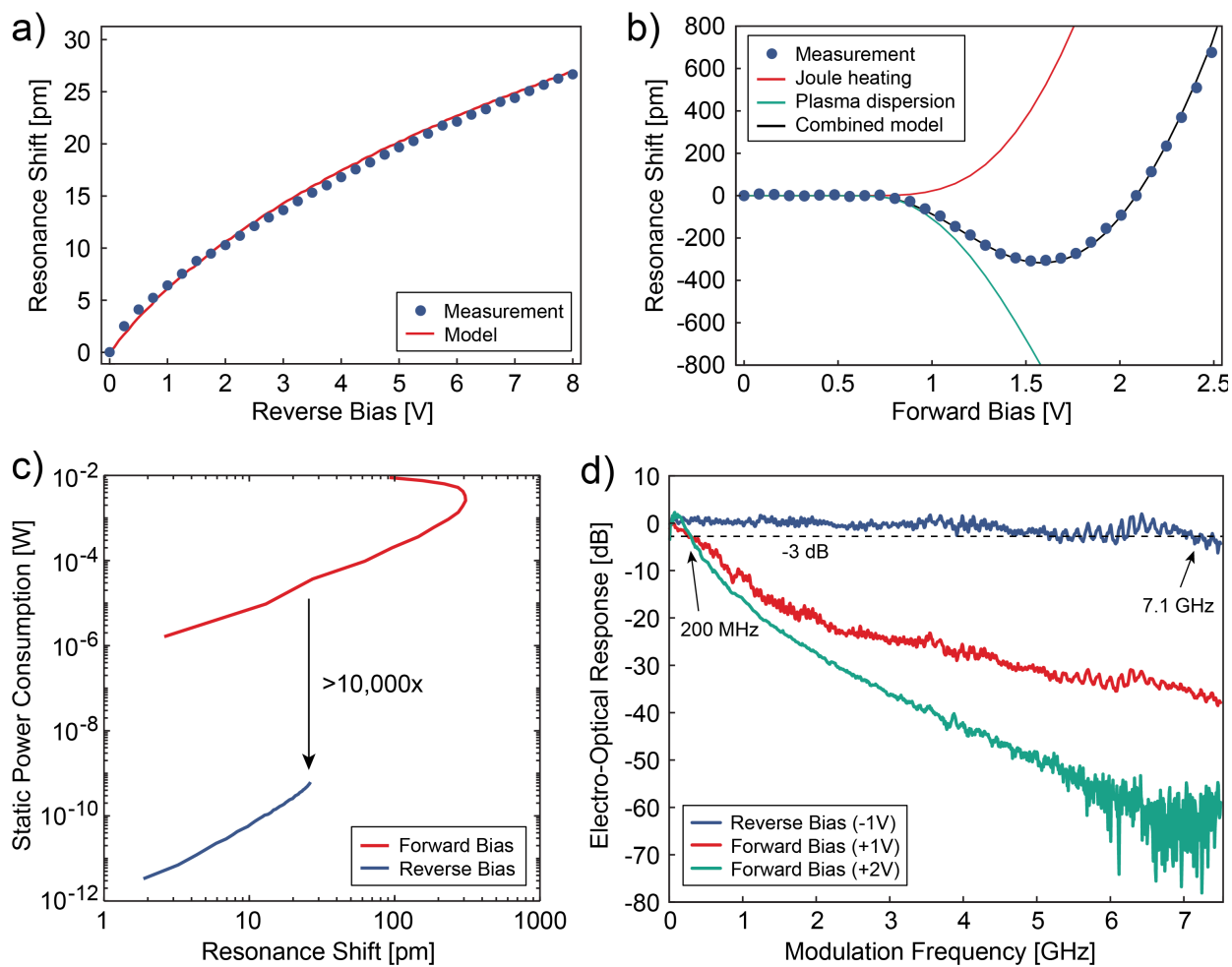

**Figure 4: Volatile tuning of photonic weight. a)** Measured resonance shift as a function of reverse bias voltage. Red line shows the expected resonance shift based on the analytical model of carrier modulation in the PN depletion region. **b)** Shift in MRR resonance as a function of forward bias voltage. After ~1.5V, Joule heating begins to dominate and shifts the resonance back to longer wavelengths. **c)** Static power consumption versus resonance shift for PN junction in both forward and reverse bias configurations. Reverse biasing the device provides >10,000× better power efficiency. **d)** Normalized frequency-dependent electro-optic response of the PN junction in both forward and reverse bias conditions. Carrier recombination limits the frequency response to ~200 MHz under forward bias, while the RF bandwidth of the measurement setup limits the response in reverse bias.

Joule heating begins to dominate, and the resonance shifts to longer wavelengths. This can be modeled using the following equation:

$$\Delta\lambda_{res} = \alpha IV + \beta I^{\gamma} \qquad (1)$$

where $V$ is the applied voltage, $I$ is the current, and $\alpha$, $\beta$ and $\gamma$ are fitting parameters. The first term in Eq. 1 corresponds to the power dissipated by the PN junction (Joule heating), while the second term is dependent on the carrier density in the waveguide (plasma dispersion). After fitting to the measured wavelength shift in **Figure 4b**, we obtain $\alpha = 156.6$ pm/mW, $\beta = -89.4$ pm/mA, and $\gamma = 0.756$. Since holes tend to dominate changes in refractive index at carrier concentrations below ~$10^{19}$ cm$^{-3}$, this value of $\gamma$ is in good agreement with the commonly used value of 0.8 from Reed et al.[32]. The contributions due to Joule heating and plasma dispersion are plotted in **Figure 4b** as red and green lines, respectively with their combined effects shown in black.

While the volatile tuning range is greater for forward biasing compared to reverse biasing, the static power dissipation is much worse. In **Figure 4c**, we compare the measured power dissipation as a function of the achievable resonance shift. While the resonance shift is limited to tens of picometers for the reverse bias case, within this tuning range, the power required to maintain this volatile state is over four orders of magnitude less than the case for forward bias. This is important when considering future scaling to large memory arrays where the static power required to fine-tune optical weights could become a source of significant energy loss for the system.

In addition to power dissipation, the maximum modulation speed under forward bias is limited by carrier recombination to several nanoseconds for a PN junction[33] and tens of nanoseconds for a PIN junction[27]. This is again orders of magnitude worse than the reverse biased case where the modulation speed is limited to tens of GHz by the RC time constant of the PN junction[22]. We compare the frequency response of our PN junction in forward and reverse bias in **Figure 4d**. A bias tee was used to provide a DC offset to the RF signal coming from a vector network analyzer with 7.5 GHz bandwidth. For the case of a positive DC offset voltage (forward biased condition), we see that the $f_{3dB}$ point is ~200 MHz, which agrees well with other forward-biased silicon PN modulators[33]. For the case of a negative DC offset, the roll-off occurs at 7.1 GHz since carriers are swept out of the PN junction under a reverse bias and carrier recombination is negligible. While this is already more than a 400× improvement compared to volatile tuning of PCM memory using a PIN junction[27], we expect the response of the memory cell to exceed 20 GHz under reverse bias based on the estimated RC time constant and photon lifetime in the ring[22]. However, in the case of **Figure 4d**, the RF bandwidth of the bias tee limits the maximum frequency response of our measurement setup.

## Conclusion

We have demonstrated a multifunctional photonic memory cell with an integrated PN microheater and modulator, functionalized with PCMs. This design allows both course tuning via the PCM and fine-tuning by reverse biasing the PN junction to enhance the resolution, programming speed, and lifetime of the memory cell. Compared to prior work[27], we improve the volatile modulation speed and tuning efficiency by more than 400× and 10,000×, respectively, while introducing a new class of waveguide-integrated microheater with low optical loss and efficient heat generation. Notably, our memory cell is compatible with commercial photonics foundry process flow with the ability to incorporate PCMs using a simple back-end-of-line post-process step. These key innovations are important for the future development of computational photonic memory arrays, which are fast, efficient, and scalable.

## Methods

### *Device Fabrication*

The devices were fabricated at Advanced Micro Foundry (AMF), Singapore, using an active silicon photonic process (AMFSiP) with oxide windows etched down to the waveguides for back-end-of-the-line materials integration. 30 nm of $Sb_2Se_3$ or GSST was deposited from single targets using an AJA Orion-3 Ultra High Vacuum Sputtering system at room temperature. The patterning of the PCM cells was performed using electron beam lithography on an Elionix ELS-G100 system with Ma–N 2403 negative resist, followed by $CF_4$ reactive-ion etching. 30 nm of $SiO_2$ was subsequently deposited via sputtering to protect the PCM from oxidation.

### *Measurement Setup:*

A tunable fiber laser (Santec TSL-550) was utilized to generate input light with wavelengths ranging from 1500 to 1630 nm for the experiments. A sixteen-channel fiber array was employed to couple light into and out of the chip via on-chip grating couplers. A two-channel custom RF Probe (S-G-S) from GGB Industries with a bandwidth of up to 40 GHz was utilized for modulation of the PN junctions. To facilitate the capture and real-time recording of optical data from the chip, the output of a fiber-coupled photodetector (Newport 2011-FC) was connected to a BNC-2110 data acquisition board from National Instruments. To capture the spectral response of the rings, the wavelength of the tunable laser source was swept from 1500 to 1630 nm at a speed of 100 nm/s, and 100,000 samples were captured from the output photodetector by the DAQ board for the entire sweep range.

For capturing the electro-optic response and bandwidth of PN microheaters, which requires operation at higher speeds, a high-speed photodetector (Newport 1544-A) with a bandwidth of up to 12 GHz and a vector network analyzer (Siglent SVA1075X) with a bandwidth between 100 KHz and 7.5 GHz were utilized. In these measurements, a bias tee (Mini-Circuits ZFBT-4R2GW+) was used to provide a DC offset to the RF signal coming from the VNA.

### *Switching Phase-Change-Materials:*

After sputtering the PCMs onto the chip, they were in an amorphous state. To crystallize the PCMs, an annealing process was used in which the chip was heated to 200C on a hot plate for 10 minutes. After initial crystallization, amorphization and re-crystallization were achieved by sending electrical pulses. For amorphization, a 100 μs pulse with amplitudes ranging from 7V to 8V was used, and for re-crystallization, a 10 ms pulse with 1 ms rise and fall times and amplitudes ranging from 4V to 5V was used. Crystallization pulses were generated using a Rigol DG4102 arbitrary waveform generator, while a 50 MHz Wavetek pulse generator was used to provide the higher voltage pulses for amorphization.

## Data availability

All data supporting this study are available in the paper and Supplementary Information. Additional data related to this paper are available from the corresponding authors upon request.

## Disclosures

The authors declare no conflicts of interest.

## Acknowledgements

This work was supported by the National Science Foundation under award numbers 2028624, 2007774, 2210168/2210169, 2329087, and 2236972, as well as AFOSR Young Investigator Award #FA9550-24-1-0064. This work was also supported by the Natural Sciences and Engineering Research Council of Canada (NSERC) and the Canada Foundation for Innovation (CFI). BJS is supported by the Canada Research Chairs program.